\documentclass[a4,11pt]{reportN}

\usepackage[brazil]{babel}      
\usepackage[utf8]{inputenc}
\usepackage{float}
\usepackage[numbers]{natbib}
\usepackage{graphics}
\usepackage{subfigure}
\usepackage{graphicx}
\usepackage{verbatim}
\usepackage{epsfig}
\usepackage[centertags]{amsmath}
\usepackage{graphicx,indentfirst,amsmath,amsfonts,amssymb,amsthm,newlfont}
\usepackage{longtable}
\usepackage{cite}
\usepackage{epstopdf}
\usepackage[usenames,dvipsnames,svgnames,table]{xcolor}

\begin{document}

\title{Análise do Circuito RLD por Espaço de Estados usando modelo Não-ideal do Diodo}

\author
    { \Large{Márcia Luciana da Costa Peixoto}\thanks{marciapeixoto93@hotmail.com} \\
   \Large{Erivelton Geraldo Nepomuceno}\thanks{nepomuceno@ufsj.edu.br} 
     \\
   \Large{Samir Angelo Milani Martins}\thanks {martins@ufsj.edu.br} \\
  {\small Programa de Pós-Graduação em Engenharia Elétrica, Departamento de Engenharia Elétrica, Grupo de Controle e Modelagem de Sistemas, UFSJ, São João del-Rei, MG} }

\criartitulo


\begin{abstract}
{\bf Resumo}. Neste trabalho é analisado o comportamento não-linear e caótico para o circuito RLD, que é modelado  por espaço de estados e simulado numericamente considerando-se como modelo do diodo a associação em 
paralelo entre resistências e capacitâncias não lineares. Assim, sendo possível caracterizar a dinâmica do circuito por meio de duplicações de período até o comportamento caótico.  Ainda calculou-se o expoente de Lyapunov para comprovar a existência de caos no circuito. E, por último comparou-se resultados simulados com experimentais garantido a qualidade do modelo obtido.

\noindent
{\bf Palavras-chave}. Circuito RLD, Modelo do Diodo, Duplicação de Período, Caos.
\end{abstract}
\section{Introdução}

\sloppy  

Sistemas dinâmicos caóticos têm recebido grande atenção nos últimos anos, 
principalmente após
Lorenz \citep{Lor1963} descobrir que seria praticamente impossível fazer previsão a longo prazo de um sistema dinâmico não-linear. 
Dentre os sistemas caóticos estudados, há circuitos elétricos como os circuitos RLD \citep{RLD}, Chua \citep{Chua}, Jerk \citep{Sprott}, conversor Buck \citep{Buck}.

Em  alguns sistemas eletro-eletrônicos, fenômenos como o aparecimento de sub-harmônicas, oscilações quase-periódicas e  comportamento caótico estão presentes, pois existe uma série de elementos que possuem relações de não linearidade entre as variáveis como: dispositivos semicondutores, capacitâncias, indutâncias e
resistências não lineares. O que ocorre muitas vezes é a linearização dos modelos matemáticos dos sistemas eletro-eletrônicos podendo ocorrer uma simplificação
de sua dinâmica, assim o modelo poderá deixar de prever tais fenômenos  \citep{Hanias2009,PCYH2017}.

Uma das marcas mais importantes do caos é a chamada sensibilidade às condições iniciais, o que significa que duas trajetórias arbitrárias inicialmente fechadas divergem exponencialmente ao longo do tempo. O expoente de Lyapunov de um sistema dinâmico é o número que caracteriza a taxa de separação dessas trajetórias infinitesimalmente próximas  \citep{Alpar2014}. Uma outra forma verificar o comportamento do sistema é por meio de atratores estranhos, que é a forma na qual os movimentos caóticos do sistema se auto-organizaram. Seu comportamento é exibido num espaço matemático abstrato, chamado espaço de fase que parece aleatório ou caótico \citep{Wheatley1996}.

Dentre os sistemas elétricos caóticos estudados, há um grande interesse em circuitos eletrônicos, devido a facilidade de implementação e análise, como os circuitos Chua \citep{Chua}, Jerk \citep{Sprott} e ainda o RLD \citep{RLD}, que é  analisado neste trabalho. O circuito RLD  é composto somente por elementos passivos conectados em série, alimentados por um sinal de tensão senoidal \citep{Hanias2009}. Para alguns determinados valores dos parâmetros as características elétricas do circuito (tensão e corrente) podem apresentar duplicação de período, caos, janelas periódicas,   histerese, entre outros efeitos \citep{Linsay1981,Rollins1982}.

Uma das formas de analisar o comportamento é por meio da modelagem do circuito em espaço de estados, que permite sua solução em \textit{softwares} como Matlab. Entretanto, para o caso do RLD deve-se ter atenção pois é necessário o modelo não-ideal do diodo. Assim, o principal  objetivo deste trabalho é a  interpretação do comportamento não-linear e caótico deste sistema, por meio da análise por espaço de estados onde é incorporado as capacitâncias e resistências não lineares ao modelo do diodo, necessárias para exibir o comportamento caótico na simulação.

Com o modelo proposto, escolheu-se a amplitude da tensão de entrada como parâmetro de controle e assim foi possível observar o circuito apresentado duplicação de período e  caos, por meio da tensão no resistor e pelo atrator estranho. Ainda, comparou-se os resultados obtidos nas simulações com resultados experimentais e considerando somente o modelo exponencial do diodo na simulação.
 E, por fim com o intuito de comprovar o comportamento caótico é calculado o expoente de Lyapunov por meio do programa Lyapmax \citep{Kodba2005}.

Na  seção \ref{sec:cp} são apresentados os conceitos básicos do circuito estudado. Na seção 3 os métodos utilizados, já na seção \ref{sec:res} os resultados são apresentados  e por fim são apresentadas as conclusões obtidas neste trabalho.

\section{Conceitos Preliminares}
\label{sec:cp}
		
\begin{figure}[ht!]
    \centering
    \includegraphics[width=6.5cm,height=3cm]{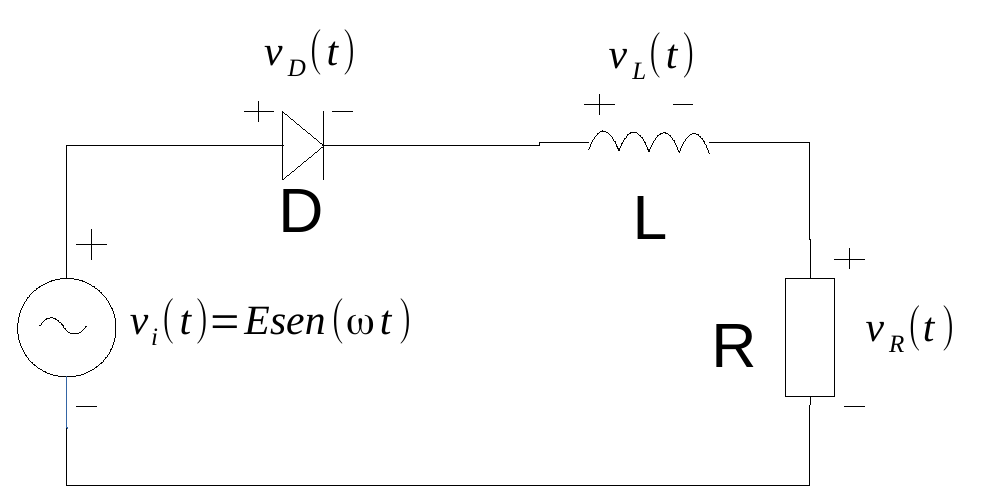}
    \caption{ \footnotesize Circuito RLD.}
    \label{fig:rld}
\end{figure}

O circuito eletrônico  RLD é composto somente por elementos passivos conectados em série, um resistor R, um indutor L e um diodo D,  alimentados por um sinal de tensão senoidal, conforme apresentado na Figura \ref{fig:rld}.
Aplicando a lei de Kirchhoff das tensões ao circuito obtêm-se a equação diferencial apresentada em (\ref{eq:1})
\begin{equation}
    Esen(\omega t) = L\frac{di(t)}{dt}+Ri(t)+v_D,
    \label{eq:1}
\end{equation}
onde  $E$  é a tensão de entrada, $v_D$ a tensão no diodo, $i(t)$ a corrente do circuito, $\omega$ é a frequência angular, $t$ é o tempo,
$D$ é o diodo,  $L$ é a  indutância e $R$ é a resistência.

\section{Metodologia}

\begin{figure}[ht!]
    \centering
\begin{minipage}{0.45\textwidth}%
\includegraphics[width=\linewidth]{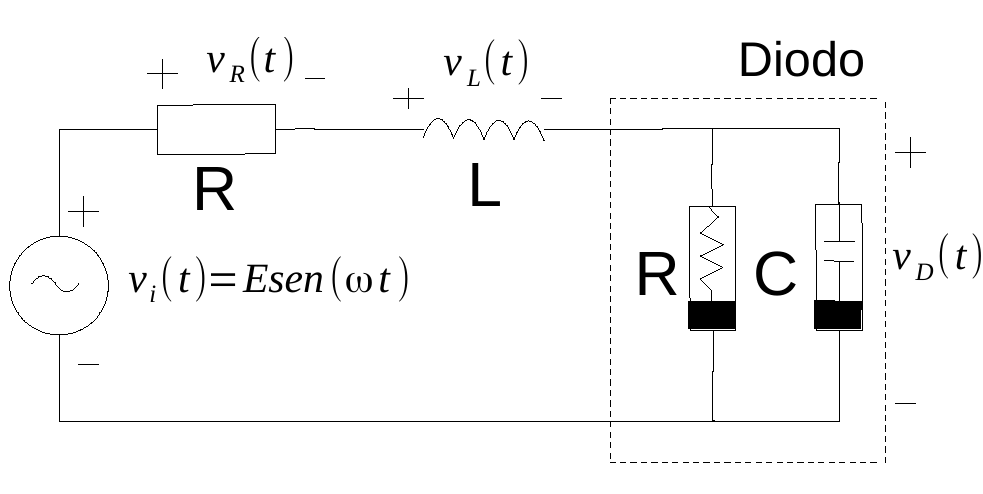} \
    \centering
    {\small (a)}%
\end{minipage}
\qquad
\begin{minipage}{0.45\textwidth}%
    \includegraphics[width=\linewidth]{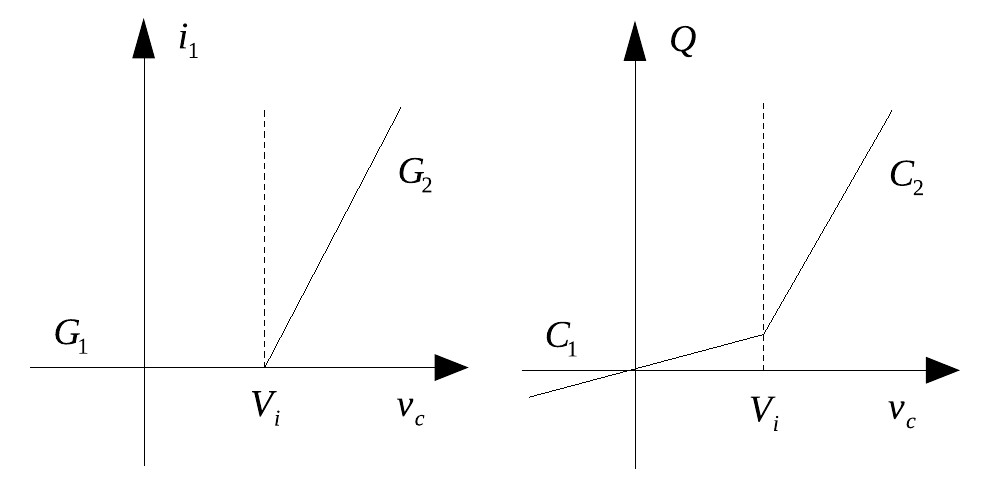}
    \centering
    {\small (b)}%
\end{minipage}
\caption{ \footnotesize (a) Circuito RLD com modelo do diodo adotado incluindo as não lineares. (b) Comportamento da resistência e das capacitâncias do modelo do diodo adotado.} \label{fig:4}
\end{figure}

A solução para o circuito, apresentado pela Equação (\ref{eq:1}), depende do modelo matemático adotado para o comportamento do diodo. Um dos modelos mais aceitos na literatura é o modelo exponencial. Entretanto, ele não incorpora as não linearidades do diodo \citep{boylestad2004}. Dessa forma, decidiu-se trabalhar com o modelo  apresentado na Figura \ref{fig:4}. Neste modelo inclui-se as capacitâncias internas não lineares, juntamente com a resistência não-linear, necessárias para exibir o comportamento caótico na simulação.
A capacitância apresentada na Figura \ref{fig:4} é a soma da capacitância de junção e difusão, o comportamento da resistência linear por partes e das capacitâncias estão representados na Figura \ref{fig:4} (b). Para tensões  maiores que a tensão de limiar, o diodo é polarizado diretamente, com  capacitância $C_2$ e consequentemente, resistência  $R_2  = 1 / G_2$, e para tensões abaixo da tensão de limiar, a capacitância é a $C_1$  e a resistência é $R_1 = 1 / G_1$ , como $ G_1 = 0$, $ R_1 \rightarrow \infty $ \citep{Matsumoto1984}.

A partir das leis de Kirchhoff para o circuito equivalente, as seguintes equações diferenciais são obtidas \citep{Azzouz1990}:
\begin{figure}[H]
    \centering
\begin{minipage}{0.38\textwidth}%
\begin{equation*}
    \frac{dq}{dt} = - \frac{G_k}{C_k}(q-q_0)+i \quad \mathrm{e} \quad  
\end{equation*}
\end{minipage}
\begin{minipage}{0.45\textwidth}%
   \begin{equation*}
    \frac{di}{dt} = -\frac{1}{LC_k}(q-q_0)-\frac{R}{L}i + \frac{Ecos\omega t - V_i}{L},
\end{equation*}
\end{minipage}
\end{figure}
\noindent onde $q$ é a carga do capacitor,  $E$  é a tensão de entrada, $V_i$ a tensão de limiar, $i$ a corrente do circuito,  $t$ é o tempo,  $L$ é a  indutância, $R$ é a resistência, $G_k$ e  $C_k$ são a condutância a capacitância, respectivamente, do modelo do diodo e os índices $k$ dependem da região de operação 1 ou 2 conforme apresentado na Figura \ref{fig:4}. Introduzindo o vetor $x= (q-q_0,i)^T$, tem-se
\begin{equation}
    \frac{dx}{dt}= A_kx+b_k(t), 
\end{equation}
onde
\begin{equation}
    A_k = \begin{bmatrix}
-\frac{G_k}{C_k} &1 \\ 
 -\frac{1}{LC_k}&-\frac{R}{L} 
\end{bmatrix} \qquad \mathrm{e}, \qquad     b_k = \begin{bmatrix}
0\\ 
\frac{E cos \omega t - V_i}{L}
\end{bmatrix}.
\end{equation}

Normalmente dispositivos eletrônicos são sensíveis a alta frequência, muitos efeitos capacitivos podem ser desconsiderados em baixas frequências, pois $X_C = 2\pi fC$ é muito grande, portanto um circuito aberto. Entretanto, em altas frequências $X_C$ ficará pequeno devido ao valor de $f$, isso  introduz um caminho de baixa reatância \citep{boylestad2004}. Como já foi visto o diodo é um elemento sensível as capacitâncias. 

Assim, para análise do circuito implementou-se a rotina em Matlab, considerando o modelo apresentado.  Nas simulações utilizou-se a tensão como parâmetro de controle e manteve-se a frequência constante em 100 KHz. Utilizou-se uma resistência de 10 $\Omega$, uma indutância de 1 mH e os parâmetros do diodo 1N4001 para a simulação do modelo. Variou-se a amplitude da senoide de entrada em 1 V, 3 V, 6 V e 9 V.

\section{Resultados}
\label{sec:res}

\begin{figure}[ht!]
    \centering
\begin{minipage}{0.39\textwidth}%
\includegraphics[width=\linewidth]{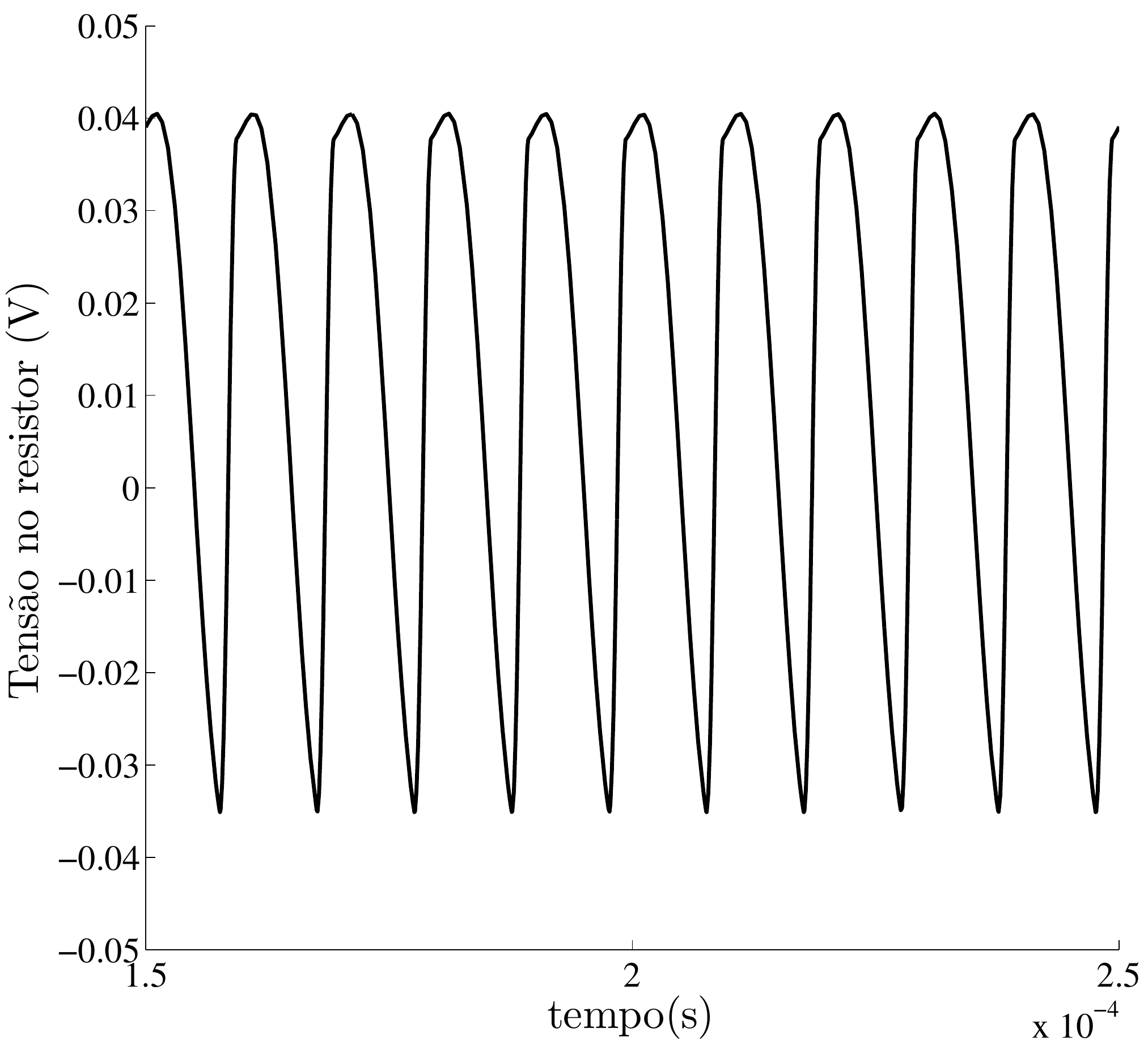}
    \centering
    {\footnotesize (a) Período 1}%
\end{minipage}
\qquad
\begin{minipage}{0.39\textwidth}%
    \includegraphics[width=\linewidth]{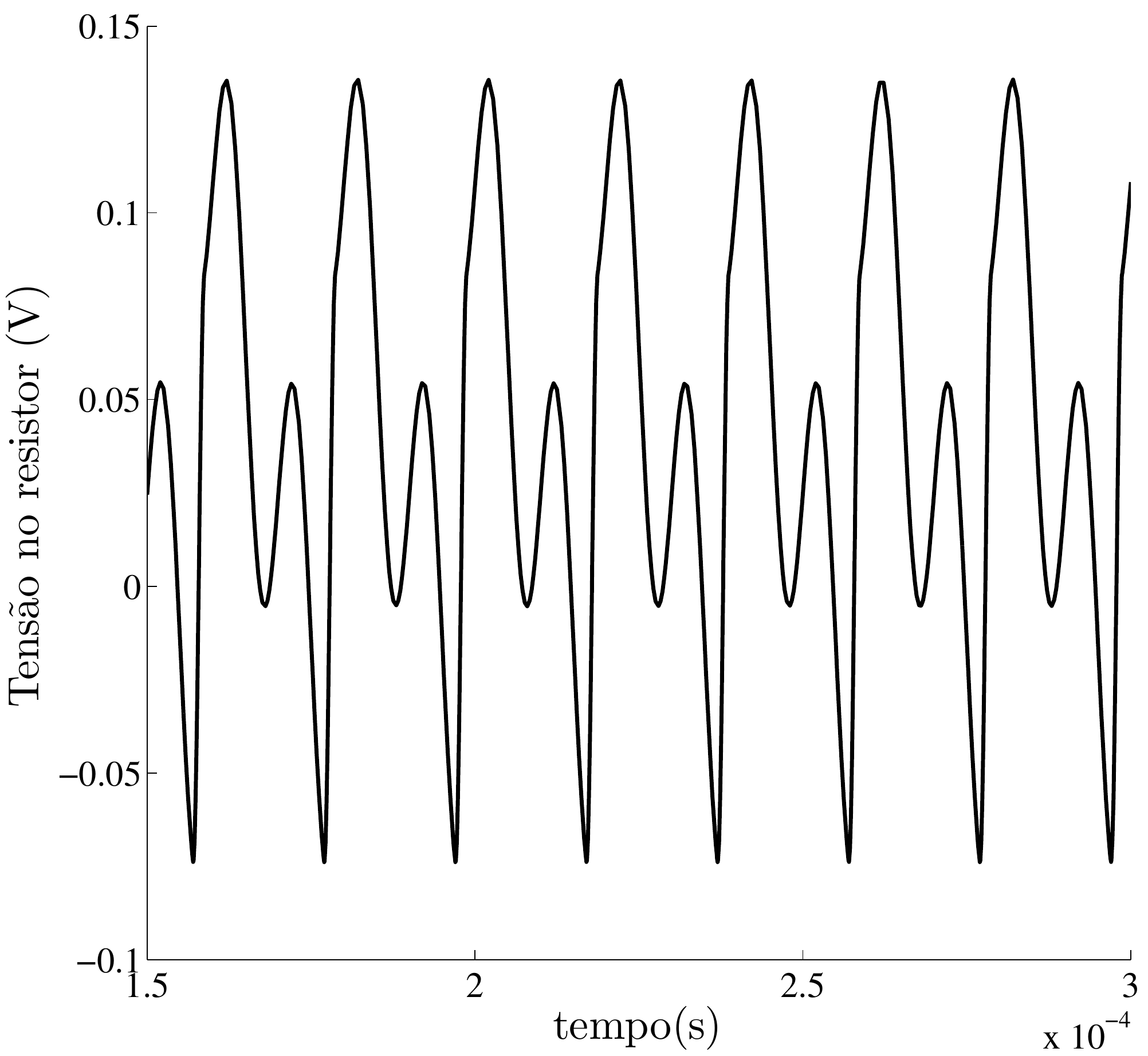}
    \centering
    {\footnotesize (b) Período 2}%
\end{minipage}
\qquad
\begin{minipage}{0.39\textwidth}%
    \includegraphics[width=\linewidth]{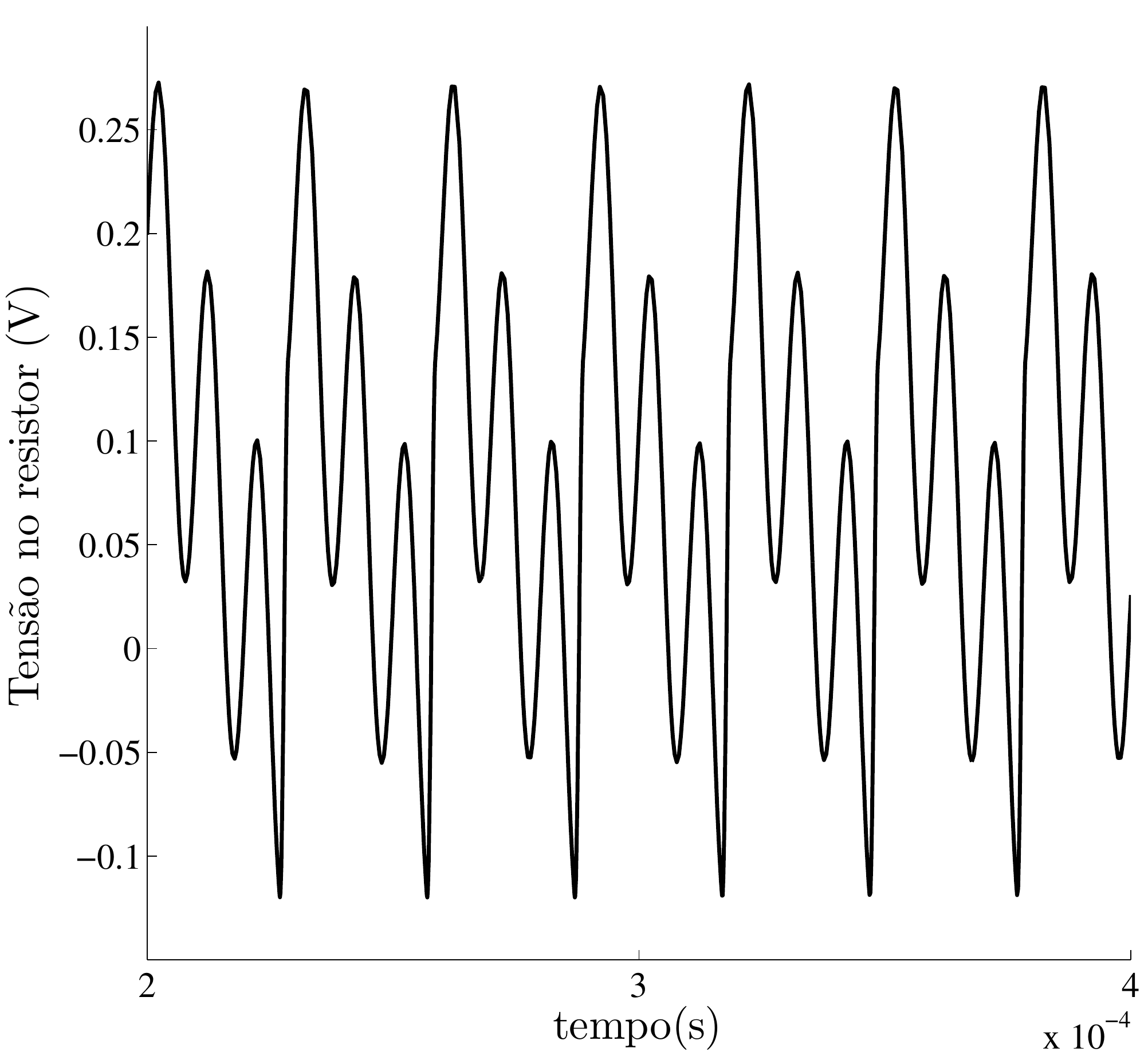}
    \centering
    {\footnotesize (c) Período 3}%
\end{minipage}
\qquad
\begin{minipage}{0.39\textwidth}%
    \includegraphics[width=\linewidth]{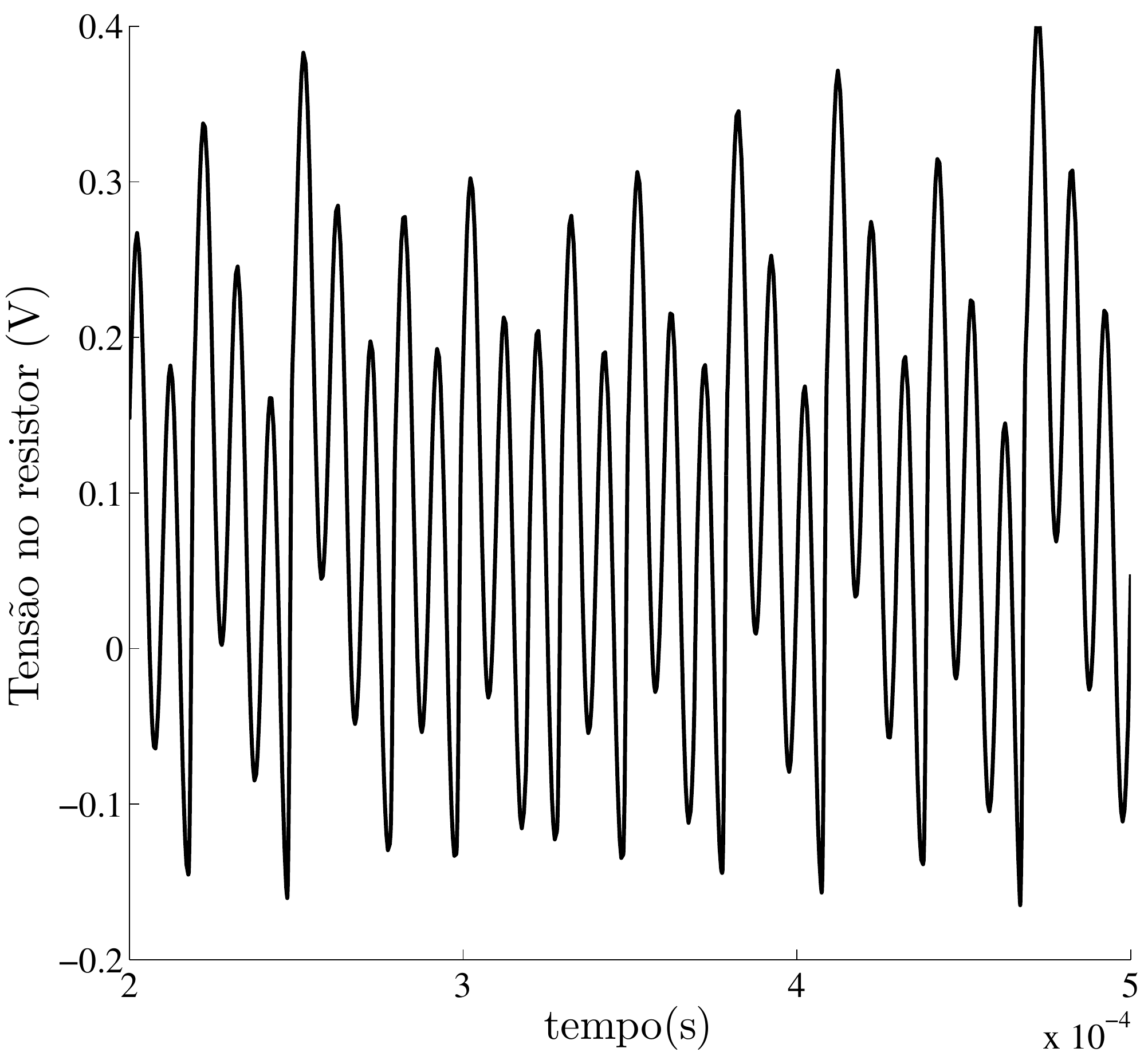}
    \centering
    {\footnotesize (d) Caos}%
\end{minipage}
\caption{\label{fig:5} \footnotesize Análise da tensão no resistor. (a) Tensão de entrada 1 V, sistema apresenta período 1. (b) Tensão de entrada 3 V, sistema apresenta período 2, (c) Tensão de entrada 6 V, sistema apresenta período 3. (d) Tensão de entrada 9 V, sistema aparentemente em caos.}
\end{figure}

\begin{figure}[ht!]
    \centering
\begin{minipage}{0.375\textwidth}%
\includegraphics[width=\linewidth]{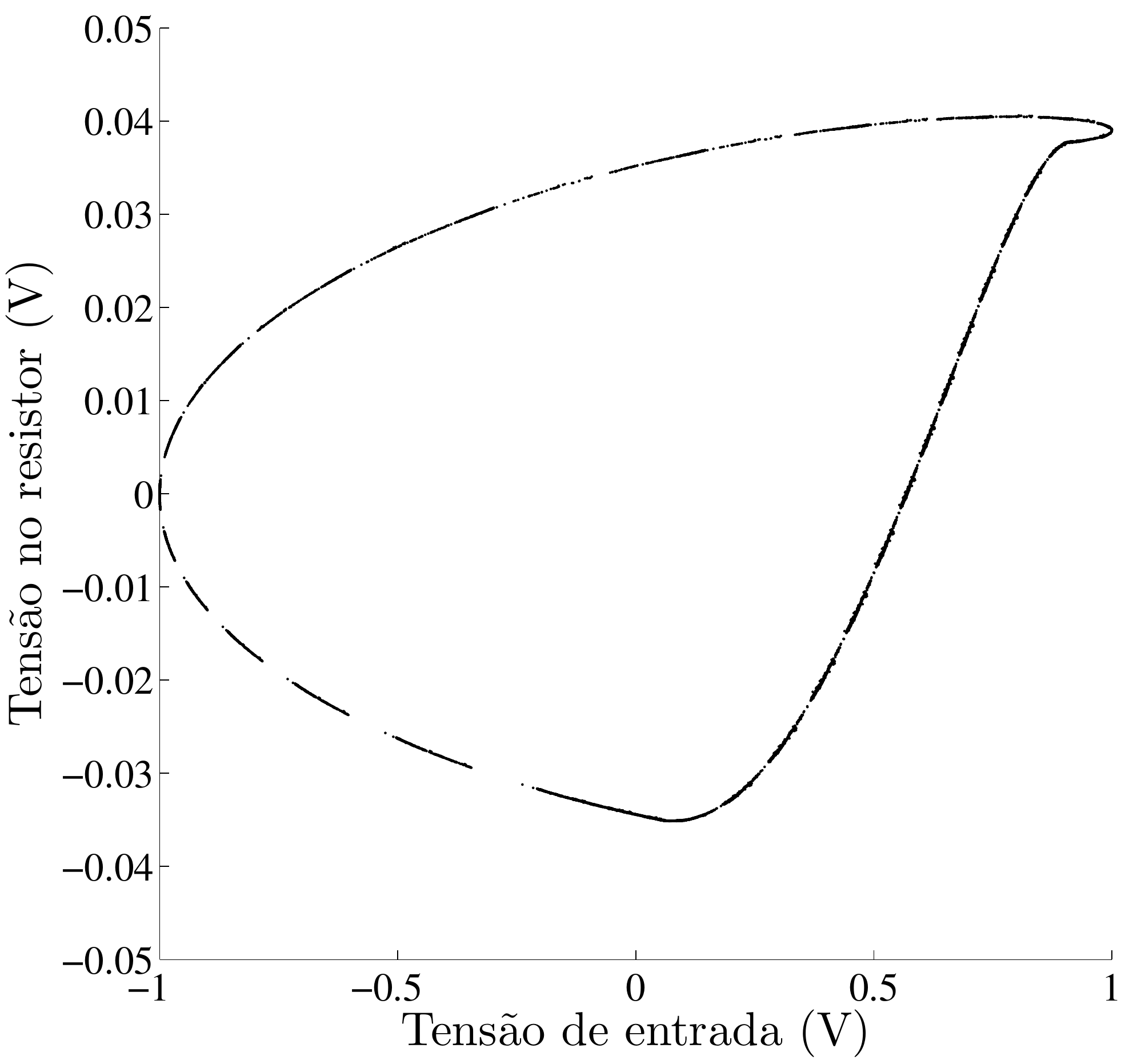}
    \centering
    {\footnotesize (a) Período 1}%
\end{minipage}
\qquad
\begin{minipage}{0.375\textwidth}%
    \includegraphics[width=\linewidth]{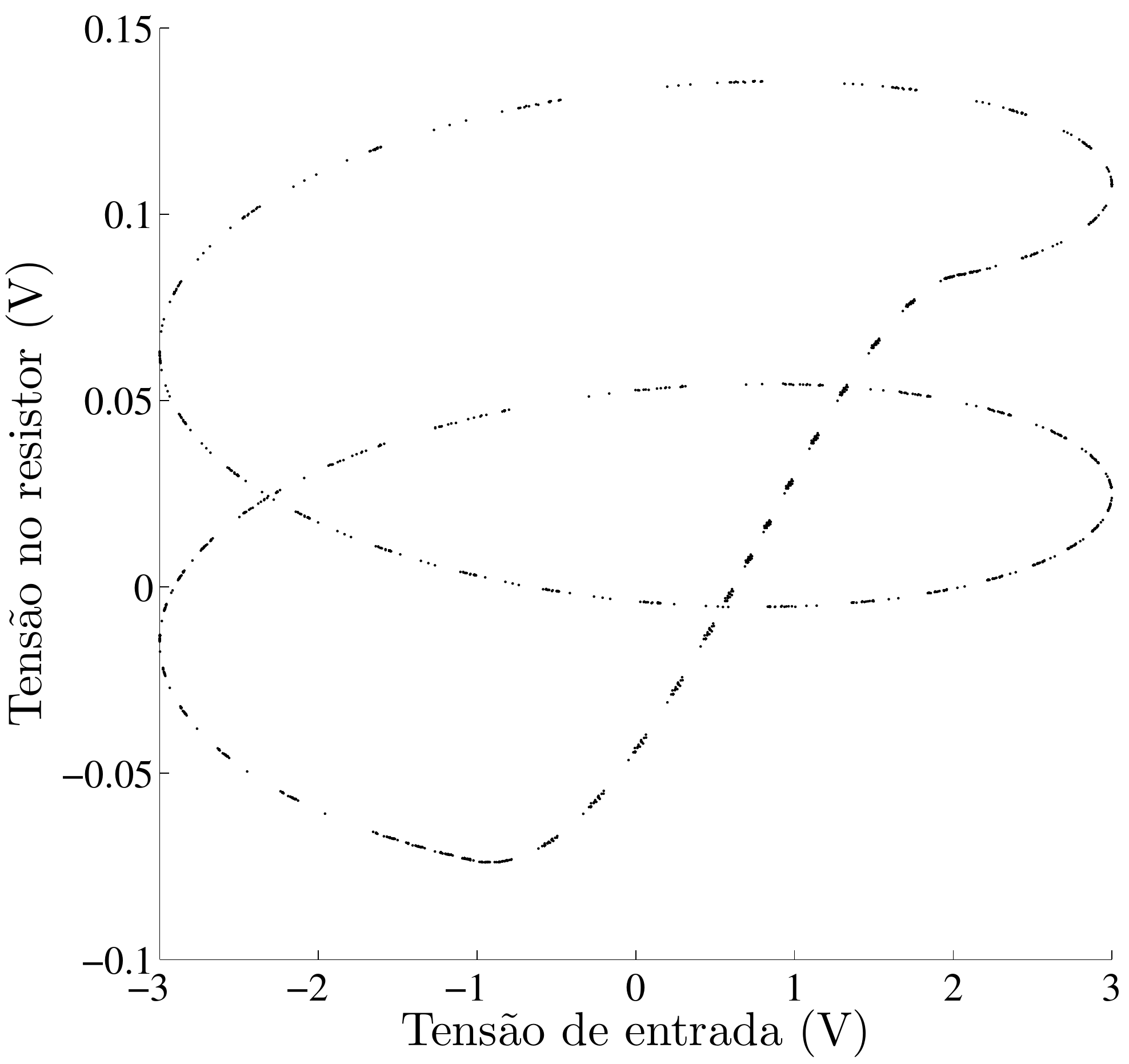}
    \centering
    {\footnotesize (b) Período 2}%
\end{minipage}
\qquad
\begin{minipage}{0.375\textwidth}%
    \includegraphics[width=\linewidth]{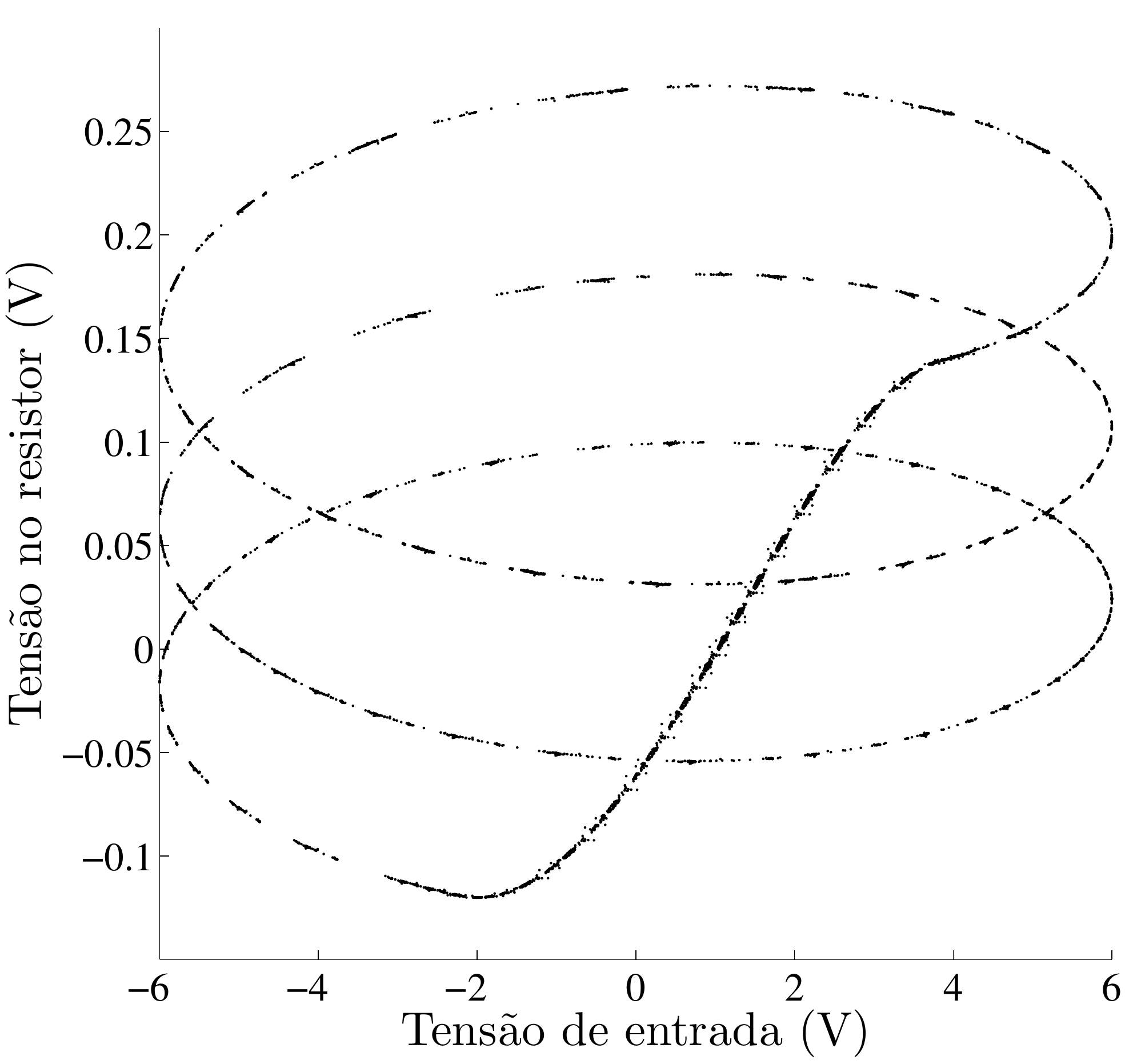}
    \centering
    {\footnotesize (c) Período 3}%
\end{minipage}
\qquad
\begin{minipage}{0.375\textwidth}%
    \includegraphics[width=\linewidth]{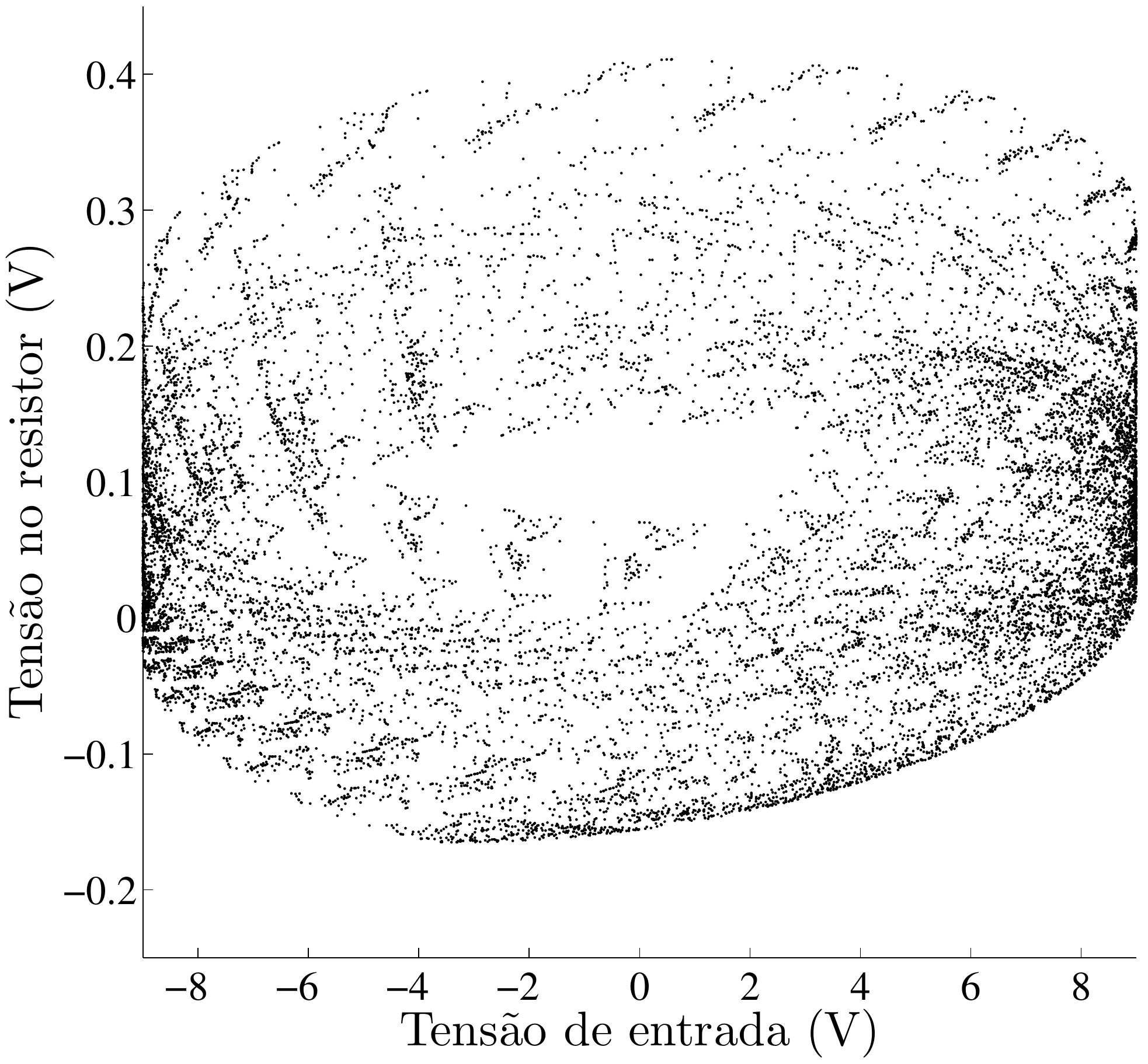}
    \centering
    {\footnotesize (d) Caos}%
\end{minipage}
\caption{\label{fig:6} \footnotesize Análise da tensão entrada versus tensão no resistor. (a) Tensão de entrada 1 V, sistema apresenta período 1. (b) Tensão de entrada 3 V, sistema apresenta período 2, (c) Tensão de entrada 6 V, sistema apresenta período 3. (d) Tensão de entrada 9 V, sistema aparentemente em caos.}
\end{figure}

A Figura \ref{fig:5}  apresenta a tensão no resistor e a  Figura \ref{fig:6}  apresenta a tensão de entrada versus a tensão de saída, ambas mantendo-se a frequência em 100 kHz e  variando a tensão de entrada entre 1 a 9 V.

\begin{figure}[ht!]
\centering
\subfigure[\footnotesize Experimental]{
\includegraphics[width=6.5cm,height=5cm]{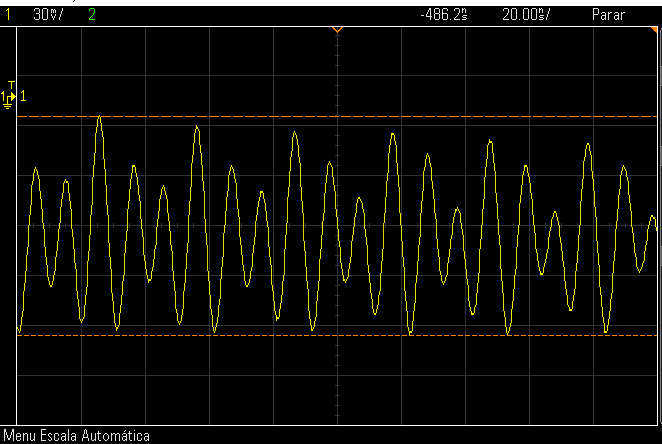}}
\subfigure[\footnotesize Modelo Exponencial]{
\includegraphics[width=6.5cm,height=5cm]{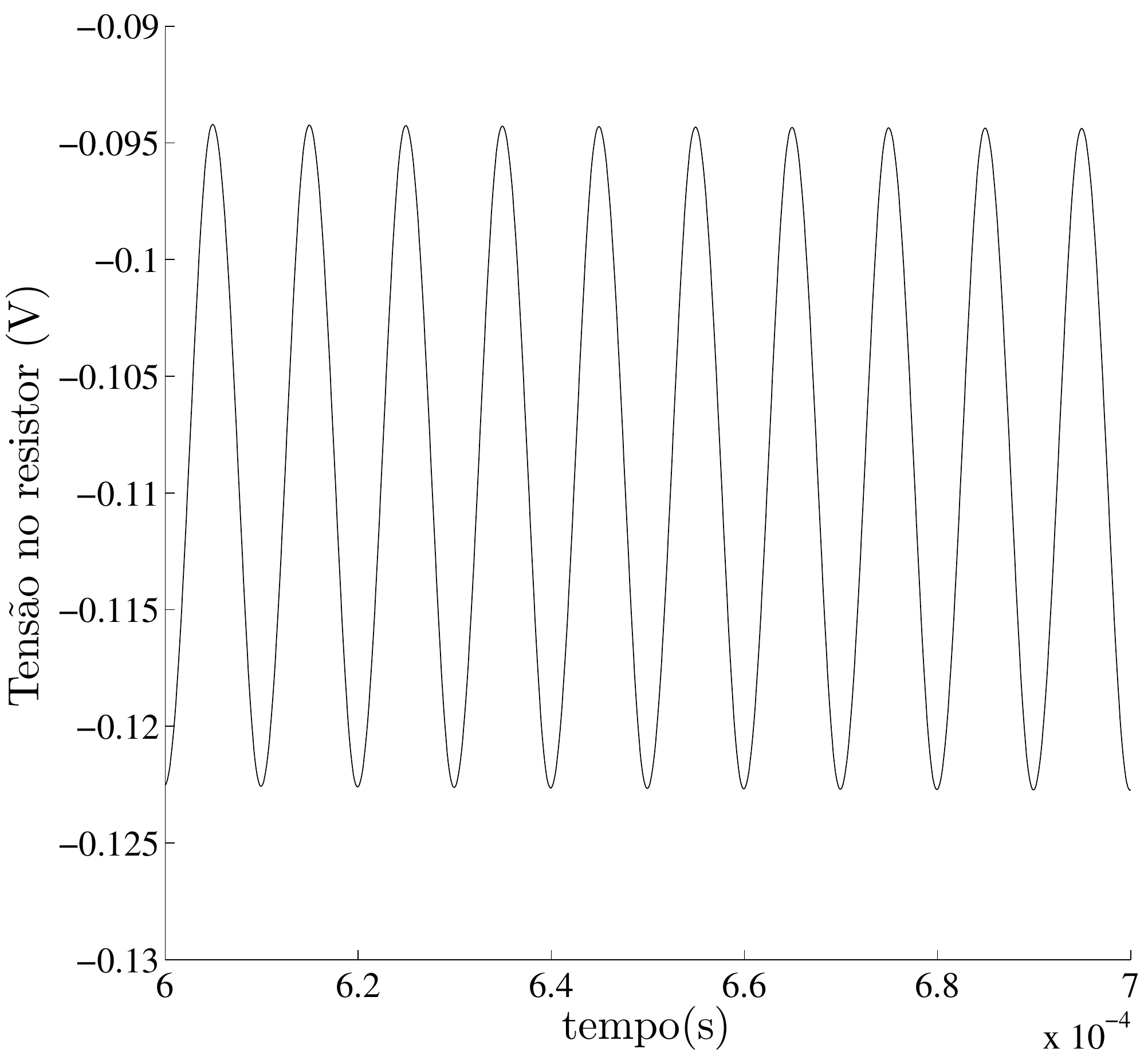}}
\caption{\label{fig:7} \footnotesize Análise da tensão no resistor, para tensão de entrada 9 V. (a) Dados experimentais obtidos por meio do osciloscópio, (b) simulação utilizando o modelo exponencial do diodo.}
\end{figure}

Para comparação dos resultados obtidos por meio do modelo apresentado, analisou-se o circuito com uma tensão de entrada de 9 V, 100 kHz de frequência, R = 10 $\Omega$ e L = 1 mH. Experimentalmente-se fez a medição utilizando um osciloscópio DSO-X, 202 A, 70 MHz. A Figura \ref{fig:7} (a) apresenta o resultado. Ainda para efeito de verificar a qualidade do modelo apresentado, considerou-se somente o modelo exponencial e simulou-se, o resultado pode ser visto na Figura \ref{fig:7} (b). Pode-se observar maior semelhança do modelo apresentado com os dados experimentais  e não com o resultado obtido utilizando o modelo exponencial, comprovando a necessidade da inserção das não linearidades ao modelo.

O expoente de Lyapunov foi calculado pelo simulador Lyapmax e o resultado obtido foi de $ 0,0770 \mu s/bits$, comprovando a existência de caos no circuito.

\section{Conclusões}
Neste trabalho analisou-se o comportamento não-linear e caótico do circuito eletrônico RLD. Para tal, considerou-se como modelo do diodo a associação em 
paralelo entre resistências e capacitâncias não lineares,
necessárias para exibir o comportamento caótico na simulação.

O modelo obtido por meio de espaço de estados para o circuito foi simulado no Matlab, utilizando-se a amplitude da tensão de entrada como parâmetro de controle e pode-se observar as duplicações de período. Ainda pode-se verificar o sistema entrando em caos, para tal constatação calculou-se o expoente de Lyapunov. E, por fim comparou-se a resposta do sistema em caos com dados experimentais e considerando-se somente o modelo exponencial para o diodo, onde se pode observar maior semelhança  do modelo apresentado com os dados experimentais  e não com o resultado obtido utilizando  o modelo exponencial, que é um dos mais aceitos na literatura, comprovando a eficiência e a necessidade da inserção das não linearidades ao modelo do diodo.

 Dessa forma, a  maior contribuição deste trabalho, foi obter uma modelagem para o circuito RLD por meio de espaço de estados que correspondesse mais precisamente ao comportamento real do mesmo.

\section*{Agradecimentos}
Agradecemos à CAPES, CNPq/INERGE, FAPEMIG e à Universidade Federal de São João del-Rei pelo apoio.

\bibliographystyle{abbrv}
\bibliography{referencias}

\end{document}